\documentstyle[11pt,newpasp,twoside,epsfig]{article}
\begin{document}

\title{The frequency content of $\delta$ Sct stars as
determined by photometry}

\author{Ennio Poretti}
\affil{Osservatorio Astronomico di Brera, Via E.~Bianchi 46, 23807 Merate, Italy}

% The abstract is entered in a LaTeX "environment", designated with paired
% \begin{abstract} -- \end{abstract} commands.  Other environments are
% identified by the name in the curly braces.

\begin{abstract}
The results obtained by means of the photometric approach to the study of
$\delta$ Sct stars are extensively discussed. The different frequency contents of
the three best candidates for asteroseismological studies (FG Vir, 4 CVn
and XX Pyx) are presented and compared; the importance of the amplitude
variations and of the combination terms is emphasized. 
The analysis of other multiperiodic variables shows how a large variety
of nonradial modes are excited; in some cases, modifications of
the power spectrum can be
observed over a few years and new modes can be seen to grow. Among monoperiodic
pulsators, constant as well as variable amplitudes can be observed. 
The difficulty of identifying an 
oscillation yielding quantum numbers is emphasized; the possibilities offered by
$\delta$ Sct stars belonging to binary systems and open clusters are
discussed. In this respect, combining the photometric and the spectroscopic
approaches could lead to a solution. A comparison is also made between low- and
high-amplitude pulsators, finding similarities.

The use of a reliable Period--Luminosity--Colour relationships toward
the shortest periods can greatly help 
mode identification in the galactic stars; moreover, it could provide an independent
verification of extragalactic distances. 

\end{abstract}

% Keywords should be included, but they are not printed in the hardcopy.


\section{Introduction}
$\delta$ Sct variables are now a well-defined class of stars.
They are located on or just above the zero-age main sequence,
in the lowest part of the classical instability strip. $\delta$
Sct stars have masses between 1.5 and 2.5 M$_{\sun}$ and they are
close to the end of the core hydrogen burning phase (Breger \&
Pamyatnykh 1998). The presence of convective
zones and related phenomena such as convective overshooting, make them
very interesting objects for the understanding of stellar evolution. 
The investigation of  pulsational properties of pre-main sequence stars
allowed their instability strip in the H-R diagram to be defined
(Marconi \& Palla 1998); some of these stars showing $\delta$ Sct variability
were discovered (Kurtz \& Muller 1999).

Photometric monitoring is the most practiced approach to study
the properties of $\delta$ Sct stars and several stars have been
deeply investigated. However, the spectroscopic approach tells us that
many modes that are not photometrically detectable are actually excited.
Rotation acts as an important factor in the increase of the number of
excited modes.

The observed frequencies are between 5 and 35~cd$^{-1}$ and 
multiperiodicity is very common; only a few stars show a monoperiodic
behaviour above the current limit of the detectable amplitude from
ground, i.e., $\sim$1~mmag. The observed modes are in the domain
of pressure ($p$) modes; there is no observational evidence that
gravity ($g$) modes are excited in $\delta$ Sct stars, even if some cases 
are suggested. At the moment, $g$-modes seem to be present only in
the $\gamma$ Dor stars, which in turn do not show $p$-modes.

The great observational effort made by several teams allows us to
handle a well-defined phenomenological scenario of the $\delta$
Sct variability. This contribution tries to summarize the results
obtained in the past years by means of extended photometric time
series. The paper is  structured as follows:
\begin{list} { } { }
\item 2.~The best candidates for asteroseismological studies
\begin{list} { } { }
\item 2.1~FG Vir: 24 independent modes
\item 2.2~XX Pyx: strong and rapid amplitude changes 
\item 2.3~4 CVn: presence of combination terms and amplitude variations
\item 2.4~Comparison between FG Vir, XX Pyx and 4 CVn
\end{list}
\item 3.~Other stars studied by the Merate Group
\begin{list} { } { }
\item 3.1~44 Tau: variable amplitude and recurrent ratio 0.77
\item 3.2~BH Psc: variable amplitude and rich pulsational content
\item 3.3~V663 Cas: growth of new modes
\item 3.4~The help of the spectroscopic approach
\end{list}
\item 4.~Monoperiodic pulsators
\item 5.~$\delta$ Sct stars in binary systems
\item 6.~$\delta$ Sct stars in open clusters
\item 7.~The frequency content of high amplitude $\delta$ Sct stars
\item 8.~Summing-up and Conclusions
\item 9.~The future: the exportation of the results on galactic stars
         to \\ \hspace*{4truemm} extragalactic research
\end{list}
 In what manner
the photometric results can be used to identify modes (i.e., to
classify the oscillation in terms of quantum numbers $n$, $\ell$ and $m$)
is discussed  by Garrido (2000). Some
improvements, both observational and theoretical, are probably
necessary to make new, substantial steps forward.
A better connection between theory and observation will allow us
to really progress in the asteroseismology of these stars.

\section{The best candidates for asteroseismological studies}
The studies and the related papers  on $\delta$ Sct stars
 follow a recurrent paradigm in the process of the
determination of the frequency content. 
In most cases a first solution, often wrongly considered a ``good'' solution, 
is obtained on the basis of a few, fragmented nights. At this stage, the
observations are hardly useful for a significant analysis.
However, since the  complicated light behaviour of our stars always
leaves some unclear facts, the same authors or another team plan a second
run. As a consequence, it appears that the solution proposed in the first
paper  was indeed preliminary.
 A very interesting result often obtained while doing this refinement
is the detection of variations in amplitude and/or in frequency for the excited
modes. Therefore, more observations are requested and, in general, they are
never sufficient \ldots\@  We can however obtain more and more satisfactory results from
the observational works on  $\delta$ Sct stars by delving deeper and deeper in this
process, even if it involves stronger and stronger efforts.
The three cases reported below are probably the best examples that the 
$\delta$ Sct community can offer.

\subsection{FG Vir: 24 independent modes} \label{fg}

FG Vir can be considered a cornerstone in the development of our knowledge
of $\delta$ Sct stars. After a few nights of observations  by L\'opez de Coca
et al. (1984), it was studied first by Mantegazza, Poretti, \& Bossi (1994)
on the basis of a single-site campaign carried out at the European Southern
Observatory, Chile. These authors proposed seven certain frequencies and a 
possible eighth one; they also claimed the presence of undetected terms,
owing to the relatively high level of noise in some parts of the power
spectrum. A successive multisite campaign (Breger et al. 1998)
confirmed the seven frequencies,
demonstrating how a single site observing run can be successful in the
detection of the main components of a multiperiodic pulsator. However, the eighth,
small-amplitude frequency was an alias  and the misidentification originated
from the combination of the spectral window with the noise. A search for the
presence of the hitherto previously undetectable terms was undertaken and the number of
frequencies increased to at least 24. However, new campaigns on this star
are considered necessary to improve the frequency resolution
and  a time baseline of several months is requested.

\begin{figure}[ht]
\setlength{\textwidth}{5.3in}
\plotone{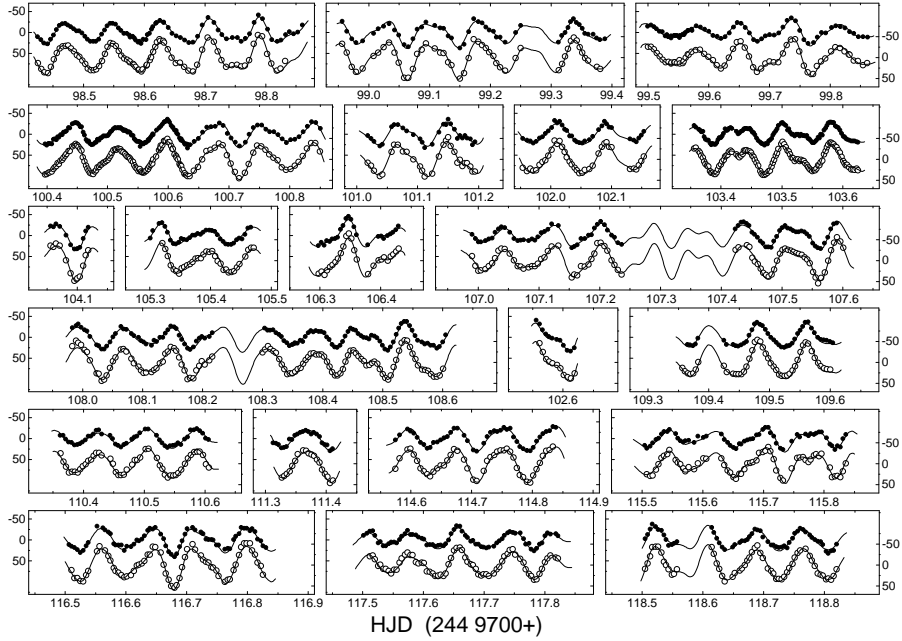}
\caption{Roughly a half of the $y$ (filled circles) and $v$ (open circles)
measurements obtained during the 1995 multisite campaign on FG Vir are shown.
The fits of the 24 frequency solutions
derived by Breger et al. (1998) are represented as solid curves}
\label{multi}
\end{figure}

The light curves obtained in a multisite campaign look quite fine and they
bear witness to the efforts made by the $\delta$ Sct researchers to
improve the quality and the quantity of the data; Fig.~\ref{multi} shows the 
very dense $v$ and $y$ light curves obtained on a baseline spanning 20 days, a subset
of the 1995 campaign. 
Considering all the frequencies now known in the light curve, it seems that
the pulsation of FG Vir is much more stable than that of  XX Pyx and 4 CVn:
the amplitude variability is very limited and the frequency values 
seem to be stable over a baseline of decades.  The frequencies are mainly 
distributed in two subgroups: the first ranges from 9.2 to 11.1~cd$^{-1}$,
the second from 19.2 to 24.2~cd$^{-1}$. An isolated peak is found at
 16.1~cd$^{-1}$ and a few between 28.1 to 34.1~cd$^{-1}$. Contrasting with
the 22 independent frequencies found, only 2 combination terms have been detected.
 It should be noted that
the identification was accepted at an amplitude S/N limit of 4.0 for an
independent term and 3.5 for a combination frequency. Kuschnig et al. (1997)
supply a theoretical basis for these assumptions. The combination terms are related
to the $f_1$ term, which has an amplitude 5 times larger than the other ones.
This term also displays an asymmetric shape, since the 2$f_1$ harmonic 
is also observed ($R_{21}=A_{\rm 2f}/A_{\rm f}=0.04$).

Photometric measurements are adequate to perform mode identification by
means of the phase shifts in different colours. Viskum et al. (1998) and
Breger et al. (1999a) agree 
that the dominant mode can be identified with $\ell$=1 and the 
12.15~cd$^{-1}$ mode with the radial fundamental.  
Considering the sophisticated pulsational modeling proposed by Breger et al.
(1999a),
FG Vir looks as a very good  candidate to match theory and observations.

\subsection{XX Pyx: strong and rapid amplitude changes} \label{xx}

\begin{figure}[t]
\setlength{\textwidth}{4.3in}
\vspace*{1.5em}
\plotone{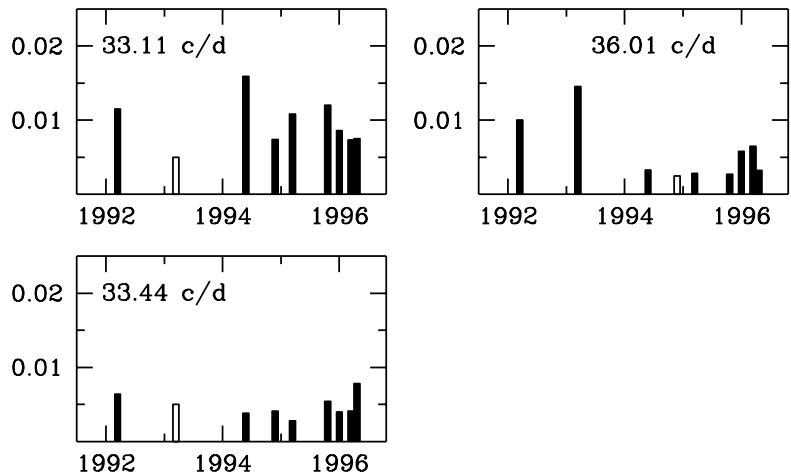}
\caption{The three main pulsation modes of XX Pyx. The rapid variability of their
 amplitude is detected by subdividing the
measurements in different subsets. White histograms indicate upper limit values
for the amplitude.}
\label{xxamp}
\end{figure}

Our knowledge on XX Pyx has rapidly  grown in the last years, following the
paradigm described above.
Its variability was discovered with the Whole
Earth Telescope (Handler et al. 1996) and then
the first solution of the light curve (based on 116.7 hours of photometry)
was already very good: 7 frequencies between 27.01 and 38.11~cd$^{-1}$ were
unambiguously found.  However, a second campaign was planned 
to match the requirements of stellar seismology. Not
only was the number of frequencies increased to 13 (Handler et al. 1997), but,
as noted above, the possibility
of comparing different observing seasons immediately evidenced 
the variability of the amplitudes. The three dominating modes change their
photometric amplitude within one month at certain times, while the amplitudes
can remain constant at other times (Handler et al. 1998): Fig.~\ref{xxamp}
shows the behaviour of these modes. The investigation about the nature of these
variations considered various hypotheses: oblique pulsator, precession of the pulsational
axis, beating of closely spaced frequencies. However, none of them 
 can explain satisfactorily the amplitude changes.
It is important to note that no change in the pulse shape of the $f_1$ mode seems
to accompany the amplitude variations, 
while changes are expected in case of frequency beating. In the
same way, evolutionary effects, binarity, magnetic field  cannot explain period changes.

The distribution of the frequencies is clearly shifted toward high values. 
A 2$f_1$ harmonic term is observed; Handler et al.
(1996) suggested the presence of very small combination terms as a representation of 
nonlinearities originating in the outer part of the star's envelope, but 
they are very close to the significance limit.

As a last step, a pulsational model of XX Pyx was undertaken, but a
unique solution could not be proposed (Pamyatnykh et al. 1998).
The presence of a frequency spacing of $\approx$$\,26\, \mu$Hz was used
to analyze different possibilities, but  
 the seismic modeling
was unable to match the observed frequencies since the mean departures exceed the
mean observational frequency by at least one order of magnitude.
Unfortunately,
this star is very faint ($V$=11.5) and line-profile variations, which could help
in the mode identifications, are very difficult to measure with the requested
accuracy. An improvement is expected from the results of a new multisite
campaign (Arentoft et al. 2000), perpetuating the observational paradigm
of $\delta$ Sct stars.
\subsection{4 CVn: presence of combinations terms and amplitude variations} \label{ai} 

Several campaigns were organized on this bright ($V$=6.1) star, allowing the
determination of a set of reliable frequencies on nine occasions over 30
years. They are mostly free from the 1~cd$^{-1}$ alias problem and are 
accurate to 0.001 cd$^{-1}$. Breger et al. (1999b) propose a
solution of the light curve composed of 18 independent frequencies and 16
combination terms $f_i\pm f_j$; the residual rms is decreased to the level
expected for the noise.

\begin{figure}[ht]
\setlength{\textwidth}{4.3in}
\plotone{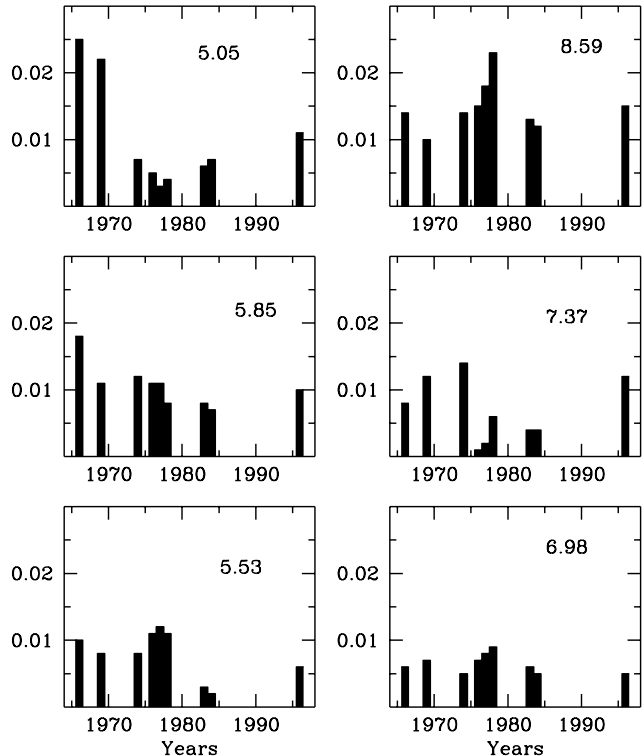}
\caption{4 CVn: the variability of the amplitude is evident for several modes.
All the available datasets are shown here}
\label{figcvn}
\end{figure}

The frequency content of 4 CVn is characterized by the grouping of all
the 18 independent frequencies in the interval 4.749--8.595 cd$^{-1}$.
In the 1996 campaign, 5 have an amplitude larger than 9 mmag; 5 others 
have an amplitude between 6.4 and 3.2 mmag. After a single peak at 1.6
mmag, the other terms (also including the combination terms) have an
amplitude smaller than 1 mmag. However, the main result of the survey
of 4 CVn is the large variability of the amplitude. This is surely not
an effect of the noise distribution, since the most relevant changes
also occur in the largest amplitude terms. Fig.~\ref{figcvn} shows 6 cases
of such variations: note
the decrease in the amplitude of the 5.05~cd$^{-1}$ term and the 
corresponding increase in that of the 8.59~cd$^{-1}$ term. Other terms,
such as the 6.98~cd$^{-1}$ term, show a more stable value for the amplitude.
It is quite impossible to discern a rule in the behaviour of the 
amplitudes. A complete discussion of the amplitude changes in the light
curves of 4 CVn is given by Breger (2000a).

The presence of a large number of combination terms is another important
aspect; they flank the set of the independent frequencies. The sums
$f_i+f_j$ are more numerous than the differences $f_i-f_j$: this should be
a selection effect since the signal at low frequencies is more difficult
to detect owing to the presence of a higher level of noise.

The pulsational modes of 4 CVn have also been investigated for the variability
of the period: some results have been obtained (Breger \& Pamyatnykh 1998), but
the matter has to be investigated further, probably considering also the 
high-amplitude $\delta$ Sct stars (Szeidl 2000). The lack of a
well-defined trend in the observed changes is the main difficulty to face
when proposing an explanation. 

%\smallskip
%\smallskip

From a methodological point of view, it is important to notice the relevant
contribution of the measurements performed by the 0.75 m Automatic Photometric
Telescope ``Wolfgang'' located at Washington Camp in Arizona, USA
(Breger \& Hiesberger 1999). As a matter of
fact, the data are of excellent quality (standard deviation 3 mmag in
$B$ and $V$ light), even if a variation in the brightness differences 
between the two comparison stars (about 2 mmag) was observed. It
coincides with the two subsets (each lasting 4 weeks) in which the
measurements are separated by a gap of two weeks. According to the
authors, the sudden variation is probably a consequence of instrumental
problems of the APT and it is not due to the variation of the comparison stars. 
The solution of this kind of problem can in the future ensure the
collection  of a large quantity of good-quality data by means of robotic
telescopes, a new frontier for the study of $\delta$ Sct stars.

\newpage

\subsection{Comparison between FG Vir, XX Pyx and 4 CVn}
\begin{figure}[ht]
\setlength{\textwidth}{4.3in}
\plotone{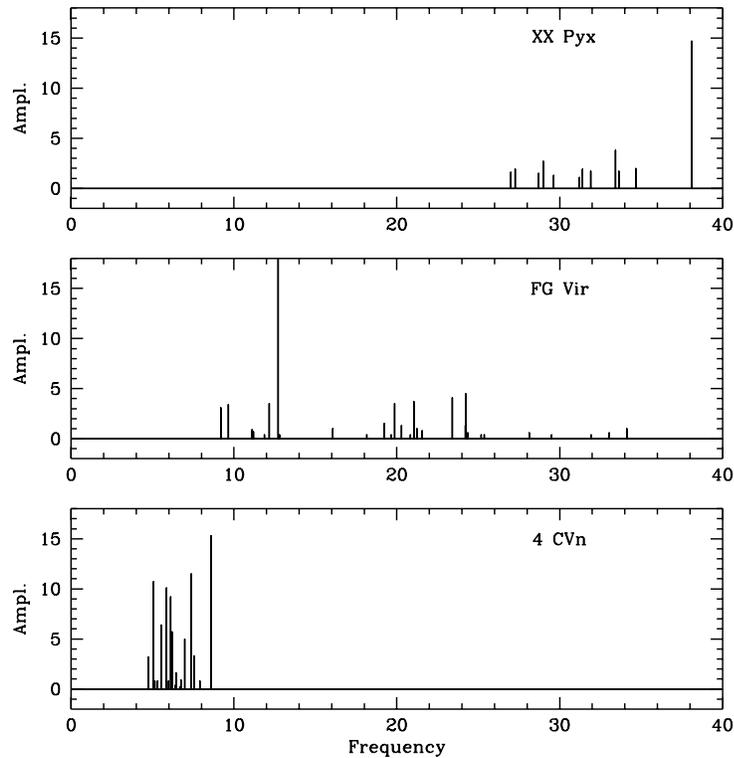}
\caption{The different distribution of the frequencies is evident in
the cases of XX Pyx, 4 CVn and FG Vir. Amplitudes are in mmag; combination
terms and harmonics are not shown.}
\label{figfre}
\end{figure}

Figure~\ref{figfre} shows the frequency content of the well-studied pulsators
described in this section; the combination terms are omitted. At first glance,
it also appears that their content is completely different: high frequencies only
for XX Pyx, low frequencies only for 4 CVn, two groups of intermediate 
frequency values for FG Vir. The obvious conclusion is that $\delta$ Sct stars are
very complicated pulsators and that a general recipe cannot be used to predict
their mode excitation.

As regards the similarities between $\delta$ Sct stars it should be noted
that the frequency content of 4 CVn matches very closely that of HD~2724
(Mantegazza \& Poretti 1999): at least 7 frequencies have almost the same
value. Among the high-amplitude terms in 4 CVn, only the 5.05~cd$^{-1}$
term has no correspondence in HD 2724. The combination frequencies, very
numerous in 4 CVn, are not seen in the light curve of HD 2724. This
can be due both to the smaller amplitude of the modes excited in HD
2724 and to the lack of a powerful tool such as a multisite campaign, not yet
exploited on HD 2724. The physical parameters of the two stars are very similar,
too.
So, it seems that we can have the possibility to group stars and not
only to observe a different behaviour for every star.

The variations of the amplitudes are a further complication, even if we can
argue that the disappearance or the damping of some terms can enhance or
make discernible other modes, increasing the number of known frequencies. The
cause of the amplitude variability is not clear: it can either be intrinsic or
originate from the beating of two close frequencies. Both in the case of
XX Pyx and 4 CVn the intrinsic damping is considered by the respective authors
as a more satisfactory explanation since the other hypothesis could not match
some observations.

\section{Other stars studied by the Merate Group}
In our effort to reveal the pulsational behaviour of
$\delta$ Sct stars the Merate group regularly performed some campaigns on selected objects.
At the beginning (second half of the eighties) there was no well-studied
object and the main goal of our studies was to monitor different variables
and try to solve the light curves as satisfactorily as possible. As a matter
of fact, the history of the worldwide observations of FG Vir started with
our campaign in 1992 (Mantegazza et al. 1994). 

Our contribution to the field was guaranteed by the extensive campaigns carried
out in Merate (as a rule one target was observed for several weeks) and at
ESO, in both cases using a 0.5-m telescope. Breger (2000b) reports on
the list of our campaigns.
Here we would like to discuss some interesting cases.

\subsection{44 Tau: variable amplitude and recurrent ratio 0.77} \label{ff}

This star has an important place in the development of the studies
performed  by the Merate group. The intensive survey carried out in 1989
allowed us to verify for the first time the variability of the amplitude
of some modes; this fact  drove a change in our approach
to the study of $\delta$ Sct  stars. It was realized that a few nights on an
object cannot constitute a significant improvement on the knowledge of the
pulsational behaviour and therefore the planning of an observational 
campaign must  satisfy critical parameters such as frequency resolution (i.e.,
a long baseline), alias damping (long nights or, better, multisite 
observations), evaluation of  observational errors and spectral window
effects, \ldots

\begin{figure}[ht]
\setlength{\textwidth}{4.3in}
\plotone{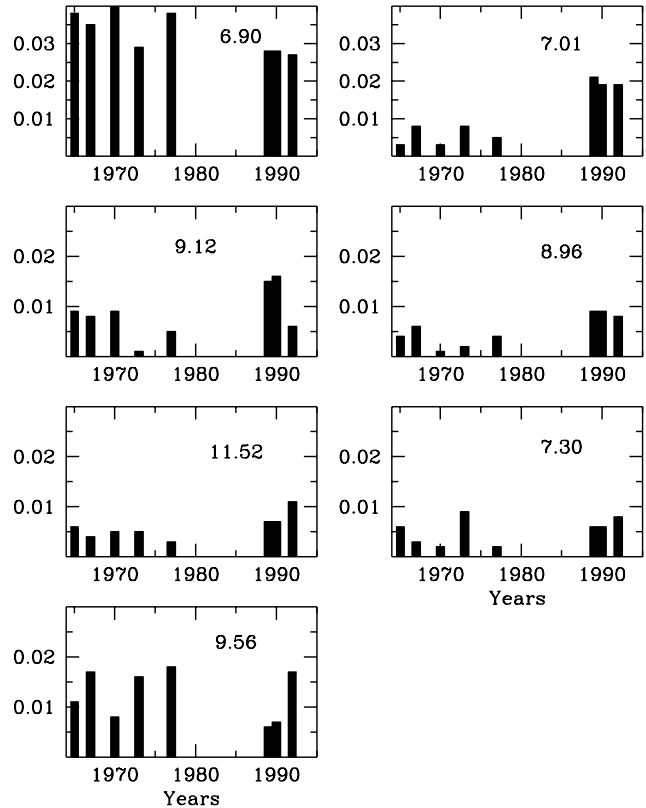}
\caption{44 Tau: the values of the amplitudes are reported for all
the available datasets and for all the 7 detected modes. Note the
opposite trend of the 7.01 and 9.56~cd$^{-1}$ terms. Amplitudes
are in mag.} 
\label{fff}
\end{figure}

Several datasets were available on 44 Tau and a comparison among them
led to the detection of amplitude variations (Poretti, Mantegazza, \&
Riboni 1992). 
However, this bright star ($V$=5.5) was considered a good target 
for other teams also; Akan (1993) and Park \& Lee (1995) provided other intensive
datasets and the verification of the amplitude variations was an important
goal of their studies. Figure~\ref{fff} summarizes the results: the
three papers quoted above supply the three values around 1990.
The 6.90~cd$^{-1}$ term is always the dominant one even if a slight decrease
has been observed in the last years, in correspondence with an increase in
the amplitude of the 7.01~cd$^{-1}$ term. Note also the similar behaviour of
the 9.12 and 9.56~cd$^{-1}$ terms and the stability of the other ones.
The general look of Fig.~\ref{fff} suggests a real variation in the
amplitudes rather than the effect of uncertainties in the amplitude 
determination.

The frequency detected in the light curve of 44 Tau also emphasizes
how complicated it is to type the modes. In several old papers the
0.77 ratio between two modes was considered as a clear fingerprint of
the pulsation in the fundamental and first overtone radial modes, as
happens in high-amplitude $\delta$ Sct stars. However, 
several examples of ratios in the range 0.76-0.78 can be found among the seven
terms: 6.90/8.96, 7.01/8.96, 6.90/9.12, 7.01/9.12, 7.30/9.56, 8.96/11.52 
\ldots \@ As a consequence, it is quite evident that nonradial modes can also
show the same ratio even in a small bunch of terms (only seven here) 
and that trying to find two frequencies showing
a 0.77 ratio to type them as radial modes is naive and misleading.

As a final remark, it should be noted that 44 Tau is probably 
a unique case in the $\delta$ Sct scenario owing to the very
small $v\sin i$ value, i.e., 4 km~s$^{-1}$. In a certain sense,
it proves that a multimode pulsation can be found in a slow 
rotator or in a pole-on star.

\subsection{BH Psc: variable amplitude and rich pulsational content}

BH Psc was observed by our group in 1989, 1991, 1994 and 1995 at the
European Southern Observatory, Chile. The analysis of the first two
campaigns showed a very complex light variability resulting from
the superposition of more than 10 pulsation modes with frequencies
between 5 and 12~cd$^{-1}$ and semi-amplitudes between 17 and 3 mmag
(Mantegazza, Poretti, \& Zerbi 1995).
The fit left a high r.m.s.\ residual 2.3 times greater than that
measured between the two comparison stars.  A second photometric campaign
was carried out in October and November 1994; we hoped to reveal more
terms and to check the stability of their amplitudes. The analysis of 
the new data allowed us to single out 13 frequencies (Mantegazza, Poretti,
\& Bossi 1996). They are concentrated
in the region 5--11.5~cd$^{-1}$: this distribution is slightly larger than that
observed for 4 CVn, but they can be considered very similar. However, more
terms should be present, with an amplitude below 3 mmag, since a good 15\%
of the variance could not be explained with the detected terms and the noise.
Moreover, the standard deviation around the mean value was considerably
higher in 1991 than in 1994 (26 mmag against 18 mmag); considerations about
the amplitudes of the modes confirmed that the pulsation energy in 1994
was lower than in 1991. BH Psc constitutes another example of a $\delta$
Sct star showing strong amplitude variations.

\begin{figure}[t]
\setlength{\textwidth}{4.8in}
\plotone{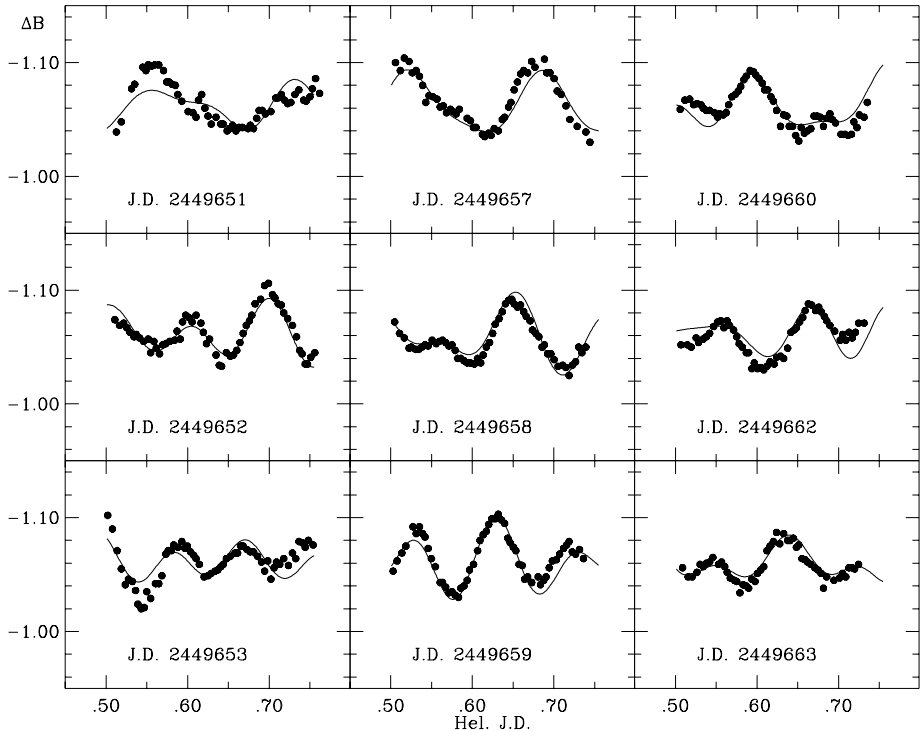}
\caption{BH Psc: $\Delta B$ differential magnitudes obtained in
1994 (dots) and their least-squares fit with the 13-sinusoid
model (solid lines): note in some parts the deviations of the points from the
calculated light curve, which suggest the presence of undetected terms.} 
\label{bhpsc}
\end{figure}

Figure~\ref{bhpsc} shows the fit of the 13-term solution to the $B$
measurements in some nights: undetected terms are the more
probable explanations for the systematic deviations which
appear on some occasions. According to the paradigm discussed in Sect.~2 and
looking at Fig.~\ref{bhpsc}, we were at the point where deeper observations
are needed and we organized 
a multisite campaign in September and October 1995.
 The analysis of the 1174 $B$ and 1251 $V$ measurements 
strengthens the hypothesis of a strong amplitude variation, probably
as rapid as in the case of XX Pyx (Mantegazza et al., in preparation).

\subsection{V663 Cas: growth of new modes} 
\begin{figure}[t]
\setlength{\textwidth}{4.3in}
\plotone{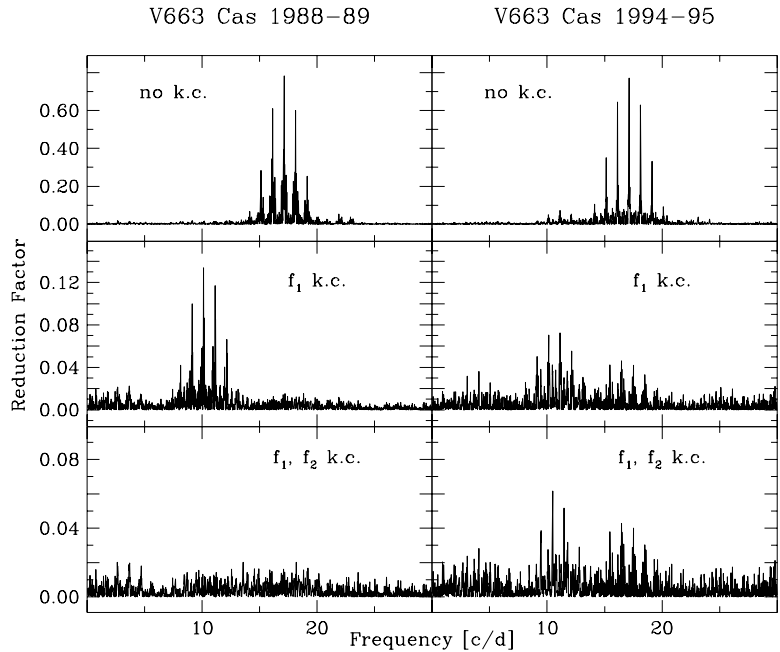}
\caption{V663 Cas: comparison between the power spectra obtained
in the analysis of the 1988--89 data (left column) and of the 1994--95
ones (right column). Each spectrum is obtained by considering the
terms detected in the previous ones as known constituents (k.c.).
The appearance of a new term is clearly visible
comparing the two bottom panels.}
\label{sao}
\end{figure}

The solution of the light curve of V663 Cas$\equiv$HD~16439$\equiv$SAO~4710
was not an
easy task. At the beginning it was considered as a monoperiodic pulsator with
$f_1$=17.13~cd$^{-1}$ 
(Sedano, Rodr\'{\i}guez, \& L\'opez de Coca 1987), but an observing 
campaign in the winter of 1988--89 
evidenced a second peak at 10.13~cd$^{-1}$ (Mantegazza \& Poretti
1990). The
amplitude of this second peak is only 3 mmag, so we considered that this
term was not detected by Sedano et al. owing to their limited dataset.
The panels in the left column of Fig.~\ref{sao} show the detection of 
these terms in the 1988--89 dataset.  The analysis of the original data
allowed us to identify the $f_1$ term (top panel). Then the frequency
of this term (but neither its amplitude or its phase) was introduced
as a known constituent (k.c.) in the subsequent iteration, in which the
$f_2$ term was detected (middle panel). Once more, these two terms were
introduced as k.c.'s: in this case no significant third term was revealed
(bottom panel).

Still following the paradigm of the study of the $\delta$ Sct stars,
a second intensive campaign was performed in the winter of 1994--95 
(Poretti, Mantegazza, \& Bossi 1996). The
results of the frequency analysis are shown in the panels of the right
column of Fig.~\ref{sao}. After the easy detection of the $f_1$ term 
(top panel), the second spectrum looks different: the structure
at about 10~cd$^{-1}$ is more complex than the one observed in the
1988--89 dataset (to compare the two middle panels). Indeed, after
considering the $f_2$ term, a new $f_3$ term was revealed at
10.48~cd$^{-1}$ and the noise around 18~cd$^{-1}$ was higher than
in the 1988--89 dataset (bottom panel).

Since the time
baseline, the number, and the standard deviation of the measurements are the
same for the two observing seasons, this fact cannot be related to a sampling
problem. Moreover, the amplitude of the $f_1$ term has slightly decreased 
from 15.9 to 14.0 mmag in $B$ light. By comparing the two observing
seasons we can discern the growth of new modes in the light curve
of V663 Cas.

\subsection{The help of the spectroscopic approach} \label {syn}

It is quite evident that every star carries a brick with which to
build our castle of knowledge about $\delta$ Sct pulsations, and
results on one star have consequences for the analysis of the next
one. However, mode identification from  photometric data always leaves
some uncertainties. For this reason we turned toward spectroscopy:
independent constraints on mode typing and detection of low and high
degree  $\ell$ modes can be obtained from the analysis of the time
series of the individual pixels defining the line profiles. Moreover,
trying to fill the gap between the output of the observations and
the inputs for a reliable pulsational model, we tried to discriminate
between different modes through a direct fit of pulsational model to
spectroscopic and photometric data. Such a {\it synergic} approach is
described by Mantegazza (2000).

\section{Monoperiodic pulsators}

It is a quite accepted concept that detecting as many frequencies as
possible in the light curves is mandatory in order to make progress
in the field of asteroseismology.  Only in this way is it possible to
compare the values predicted by the theoretical models with the observed
ones. Consequently, the observation of monoperiodic stars has been put
aside in the last decade. However, even these variables display a wide
variety of intriguing behaviour.  

\subsubsection{Beta Cas:} It is probably the best known monoperiodic
pulsator, also included in the secondary target list of the MONS
satellite. Riboni, Poretti, \& Galli (1994) proved that a constant value
of the frequency cannot fit the times of maximum brightness collected
since 1965; on the other hand, a constant value (9.897396~cd$^{-1}$)
can fit all the datasets since 1983. Hence, a probable variation of the
period occurred between 1965 and 1983. On the other hand, the amplitude
has remained stable ($\Delta V$~=~0.03 mag) over a 25-year baseline since
no significant variation could be inferred.  Considering the value of
the period (0.101 d), this behaviour is typical for a high-amplitude
$\delta$ Sct star rather than for a small amplitude one.

\subsubsection{\it AZ CMi:} The light curve of AZ CMi is asymmetrical
($R_{21}$=0.13 in $V$ light; Poretti 2000) and very stable in shape and
amplitude  ($\Delta V$=0.055 mag); hints of a possible decreasing value
of the amplitude need more observations to be confirmed. $B$ and $V$
photometry is available: the measured phase shift is $\phi_{B-V}-\phi_V$=
--9$\pm$9 degrees. Since it is negative, such a value slightly suggests
a nonradial mode, but the error bar hampers a definite identification
(Poretti et al. 1996).

\smallskip

The examples reported above seem to suggest that monoperiodic stars are
simple and stable pulsators even if the ambiguity
between radial or nonradial modes cannot be solved. However, the analysis of 
other cases complicates the scenario a little:
\smallskip

\subsubsection{\it BF Phe:} All the measurements available on this star
are well satisfied by the single term $f$=16.0166 ~cd$^{-1}$ (Poretti et
al. 1996).  The reality of a secondary peak was discussed and, at the end,
rejected. The amplitude observed in 1991 in $V$ light is very similar to
that observed in 1993 in $y$ light. However, in $b$ light the amplitude
observed in 1989 (25 mmag) is larger than that observed in 1993 (16 mmag)
and this discrepancy cannot be explained by errors in the measurements;
unfortunately  BF Phe was measured in 1989 in the $b$-light only. Hence,
this monoperiodic pulsator shows an amplitude that has changed by a
factor of 1.7 in two years, if we consider the amplitude was the same
in 1991 and 1993.

\subsubsection{\it 28 And:} Rodr\'\i guez et al. (1993) demonstrated
that 28 And$\equiv$GN And is  a monoperiodic pulsator. They carried out
a frequency analysis on the datasets available in the literature and when
the main frequency is prewhitened the resulting periodograms do not show
any trace of another significant peak. New $uvby$ photometric data were
acquired in 1996 (Rodr\'\i guez et al. 1998); the observed amplitude
of the light curves is very small  compared with any other previous
dataset. More precisely, the amplitude was about 19 times smaller than
that observed five years before. In this new dataset a second term was
detected; however, the reality of this term is not well-established,
since it is marginally significant (S/N=4.1 and other smaller peaks are
visible in the power spectrum). Further observations are requested to
verify if 28 And will display a pulsational behaviour similar to that
of V663 Cas, with new frequencies slowly growing to a detectable level.

The cases of genuine monoperiodic $\delta$ Sct stars as  BF Phe and
(probably) 28~And yield the observational evidence that in
these pulsators also the amplitudes of the excited modes are affected
by variations. Moreover, a multiperiodic pulsator as V663 Cas can show 
a different pulsational content in different seasons, greatly complicating
observational investigation. 
As a matter of fact we proved that the changing amplitude is not
a prerogative of pulsators with a large number of excited modes (as observed
in XX Pyx, 4 CVn and 44 Tau). It is also possible to find a $\delta$ Sct
pulsator displaying a single period with a very small amplitude:
the light curve of HD~19279 has a full amplitude of only 4.2 mmag
(Mantegazza \& Poretti 1993).

Moreover, the detection of a single frequency in the light curve is not
an indication of radial pulsation for $\delta$ Sct stars of low amplitude.
Table \ref{tabmono} lists the physical parameters as determined
 by means of Hipparcos parallaxes and Str\"omgren
photometry: as can be noticed, these stars display a large variety of 
pulsational constants $Q$'s and, hence, different excited modes.
\vspace*{-1.0em}

\begin{table}[b!]
\caption{Determination of the pulsational constant $Q$ for the monoperiodic
$\delta$ Sct stars discussed here. M$_{\rm bol}$ values were determined by
using the Hipparcos parallaxes} \label {tabmono}
\begin{center}
%\scriptsize
\small
\begin{tabular}{llr c r r r r}
\tableline
  \multicolumn{2}{c}{Star} & HIP & & \multicolumn{1}{c}{T$_{\rm eff}$} &
  \multicolumn{1}{c}{$\log$ g} & \multicolumn{1}{c}{M$_{\rm bol}$} &
  \multicolumn{1}{c}{$Q$} \\
  \multicolumn{2}{c}{ } & &  & \multicolumn{1}{c}{[K]} &
  \multicolumn{1}{c}{} & \multicolumn{1}{c}{} &
  \multicolumn{1}{c}{[d$^{-1}$]} \\
%\hline
\tableline  
\noalign{\smallskip}
AZ      & CMi & 37705 & & 7800 & 3.6 & 0.8 & 0.020 \\
BF      & Phe & 117515& & 7500 & 4.1 & 2.5 & 0.034 \\
28      & And & 2355 & & 7500 & 3.7 & 1.4 & 0.018 \\
$\beta$ & Cas & 746 & & 7000 & 3.6 & 1.2 & 0.020 \\
%$\tau$  & Peg & & 8180 & 3.8 & 1.0 & 0.016 \\
\noalign{\smallskip}
\tableline
\end{tabular}
\end{center}
\end{table}

\section{$\delta$ Sct stars in binary systems}
In principle, the study of $\delta$ Sct stars belonging to binary
systems can supply useful suggestions to understand the physics
of the pulsation driving. The complications introduced by the 
companion are much less important than the measurable parameters we can
attain. Lampens \& Boffin (2000) provide a review on
$\delta$ Sct stars in stellar systems. Here we would like 
to draw attention to a few binary systems:
\smallskip

\subsubsection{$\theta^2$ {\it Tau}:} It is the component of a wide binary
($P$=140.728~d) and there is no interaction between the two
stars. However, it is a well-studied pulsator. Breger et al. (1989)
determined five frequencies, very close to each other (from 13.23 to
14.62~cd$^{-1}$), without regular spacing among them. Such a  very close
(much closer than in the case of 4 CVn), isolated multiplet is not usual
in the solution of light curves. The amplitudes of the modes seem to be
very stable, even if recently Zhiping, Aiying, \& Dawei (1997) claimed
to have evidence of amplitude variations; further campaigns could be useful
to clarify the matter.

\subsubsection{$\theta$ {\it Tuc:}} This double line spectroscopic
binary is located near the South Celestial Pole, which makes it very
suitable for long-term monitoring. The orbital period is 7.1036 d;
a regular spacing is observed among the 10 detected frequencies, which
take place in the interval 15.8--20.3~cd$^{-1}$ (Papar\'o et al. 1996).
Low-frequency terms have been observed and they are ascribed to the
orbital period.  The mode identification was performed by Sterken (1997)
using the Str\"omgren photometry:  the results are not conclusive for
most of the frequencies, but the $f$=20.28~cd$^{-1}$ is reconcilable with
a radial mode.  De Mey, Daems, \& Sterken (1998) confirmed this result
and they also found that the system has a very low mass ratio, $q$=0.09.
The possibility that the regular spacing is related to the slow rotation
of the $\delta$ Sct star deserves further attention. De Mey et al. (1998)
also found 1.7$\,$$<R_1$$\,$$<2.7\, R_{\sun}$ for the radius $R_1$  of the
$\delta$ Sct star. A rotational period of 7.1036~d is not plausible since
it should imply 12$\,$$< v_1$$\,$$<19$~km~s$^{-1}$ for the rotational
velocity $v_1$, a value irreconcilable with the observed $v_1 \sin i$:
35~km~s$^{-1}$ (Mantegazza 2000); even higher values are reported in
the literature, see Sterken (1997).

\subsubsection{{\it AB Cas}:} 
\begin{figure}[ht]
\setlength{\textwidth}{4.3in}
\plotone{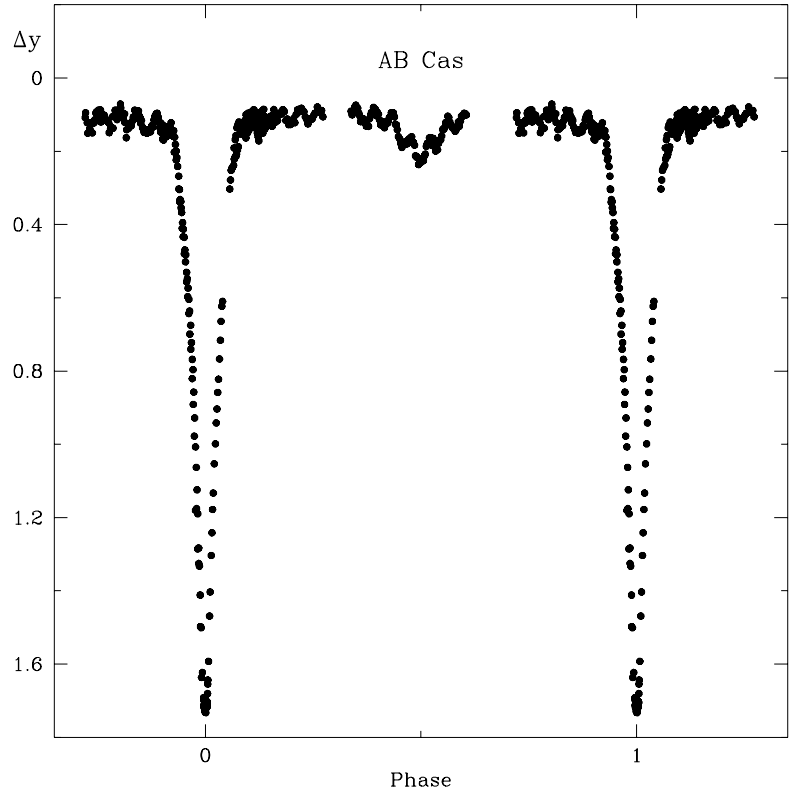}
\caption{The pulsation of the primary component of AB Cas is superimposed
on the light curve caused by eclipses}
\label{ab}
\end{figure}

AB Cas is an eclipsing binary with $P_{\it
orb}$=1.367 d; the primary component is also a monoperiodic pulsator with
$P_{\it puls}$=0.058 d (Fig.\ref{ab}); there is no connection between the
two periods (Rodr\'{\i}guez et al. 1998).  The light and colour curves
are quite normal for this variable: perhaps its main peculiarity is
its monoperiodicity.  It should be noted that the pulsational period is
among the shortest observed in $\delta$ Sct stars. The pulsation seems
to disappear when the secondary star transits across the disk of the
primary: this is due to the large depth of the primary eclipse. However,
when the curve due to binarity is removed, the pulsation is visible in
the residuals. In fact, during the primary eclipse we can see about 20\%
of the disk of the pulsating star. Unfortunately, the observational
uncertainties do not allow one to discriminate between radial and
nonradial modes by using the changes in the amplitude caused by the
progress of the eclipse. Other considerations suggest a fundamental
radial mode.

\subsubsection{{\it 57 Tau}:} Papar\'o et al. (2000) indicated 57 Tau as a 
pulsator showing simultaneously gravity and pressure modes.
 However, the discovery of its duplicity (Kaye
1999) explains the low-frequency terms as due to the orbital effect
($P_{\it orb}$=2.4860 d); the first example of a pulsator combining the
$\gamma$ Dor (gravity) and $\delta$ Sct (pressure) modes has so far not been
found. The promising case of BI CMi (Mantegazza  \&  Poretti 1994) 
is still awaiting closer investigation.
\smallskip
\smallskip

As a concluding remark on this subject, the connection between
pulsation and duplicity does not seem to be well-exploited in current
research on $\delta$ Sct stars. It is quite obvious that simultaneous
photometry and spectroscopy of a pulsator in an eclipsing system  can
help mode typing. One of the main difficulties met in the combined
use of photometric and  spectroscopic data on $\delta$ Sct stars (the
synergic approach described in Sect.~\ref{syn}) is to put constraints
on the inclination angle. The possibility of determining this angle and
the rotational period (as obtained from the orbital solution, assuming
synchronous rotation and equatorial orbits) greatly simplifies mode
identification by reducing the number of admissible modes.  Lampens \&
Boffin (2000) report on many other candidates to study the connection
between duplicity and pulsation.

\section{$\delta$ Sct stars in open clusters}
$\delta$ Sct stars belonging to an open cluster offer a very good
opportunity to delve deep into the models thanks to the more precise
information available on metallicity, distance and age. The common
origin of all the stars allows closer comparisons  between the
results found on each of them. Interesting results were found by Hern\'andez
(1998) by analyzing the data collected by the STEPHI observational
network on $\delta$ Sct stars located in the Praesepe cluster.
Unfortunately all  these stars are very complicated pulsators and 
it is very difficult to define a clear picture of their frequency
content.

Alvarez et al. (1998) discussed a set of frequencies for two of
them, BQ and BW Cnc. In particular, BW Cnc displays two pairs of very
close frequencies; inspection of the light curves showed how two close
frequencies can disappear or be misidentified in the presence of bad time
resolution. These authors recommend long time baselines for campaigns
on $\delta$ Sct stars, i.e., much longer than 1 week. Also considering
the results described below, this requirement has to be considered
as mandatory for further works. BQ Cnc is a binary system, but only
three frequencies were identified; one of them could be a $g$-mode
and further investigations are requested.  Interesting results on the
Praesepe pulsating stars were also obtained by Arentoft et al. (1998),
who successfully used defocussed CCD images. The number of detected
frequencies was limited, but the possibility of using  the Praesepe
metallicity leads the analysis toward deriving reliable physical stellar
parameters.  Pe\~{n}a et al. (1998) also re-analyzed photometric time
series of $\delta$ Sct stars in Praesepe proposing mode identifications on
the basis of the physical properties of the cluster.  Unfortunately, the
uncertainties due to the effects of rotation and convective overshooting
limit these conclusions somewhat.

The STACC network monitored some northern open clusters
(NGC~7245, NGC~7062, NGC~7226, NGC~7654) searching for $\delta$
 Sct stars (Viskum et al. 1997). As a result, they found
that the fraction of these variables is much lower than
among field stars and in other open clusters. This 
observational fact suggests that some parameters are
working in the selection of the pulsation excitation
and future efforts should be undertaken to discover
what they are.

\section{The frequency content of high amplitude $\delta$ Sct stars}

After examining the complex phenomenological scenario of low amplitude
$\delta$ Sct stars, it is interesting to verify what happens in the domain
of the high amplitude $\delta$ Sct stars (HADS). Rodr\'\i guez et al. (1996)
analyzed the multicolour data (Str\"omgren and Johnson systems) for
monoperiodic HADS. The results indicate that all these stars, both
Pop.~I and Pop.~II objects, are fundamental radial pulsators. This
conclusion was obtained on the basis of the phase shifts and amplitude
ratios between light and colour variations.

Moreover, Rodr\'{\i}guez (1999) analyzed all the reliable photometric
datasets of a selected sample of monoperiodic HADS to study the
stability of the light curves. In this manner more than 22000
measurements were scrutinized for seven stars (ZZ Mic, EH Lib,
BE Lyn, YZ Boo, SZ Lyn, AD CMi, DY Her). The conclusion was that no
significant long-term changes of amplitude occurred for any of
these stars.
\smallskip

The impression is that the HADS are very stable fundamental radial
pulsators, with a few exceptions represented by double-mode HADS.
Can such an idyllic scenario resist the observational effort made in the
last few years? It seems it does, since a large homogeneity was found
by Morgan, Simet, \& Bargenquast (1998) analyzing the HADS contained
in the OGLE database. We note here that the subgroup with an anomalous
$\phi_{31}$ parameter seems to be an artifact (Poretti, in preparation).

However, we can single out a few interesting cases among the HADS.

%\smallskip

\subsubsection{V1162 Ori:} Arentoft \& Sterken (2000) discuss the time
series collected on V1162 Ori. After a period break and a  significant
decrease in amplitude (about 50\%; Hintz, Joner, \& Kim 1998), they
also found that the period was no longer valid in early 1998 and that
the period changed again during March--April 1998. The latter change
was accompanied by an increase in amplitude of the order of 10\%; such a
phenomenology calls to mind the behaviour of some low-amplitude $\delta$~Sct
stars. A new campaign is in progress to clarify the reasons for
these changes.

%\smallskip

\subsubsection{\it V974 Oph:}
The case of V974 Oph is more intriguing. This faint ($V$=11.8) variable
was observed twice at ESO (July 1987 and April 1989). It was first
considered a HADS similar to V1719 Cyg (Poretti \& Antonello 1988), but
the second run clearly demonstrated that it is a multiperiodic star:
at least 4 independent frequencies were detected ranging from 5.23 to
6.66~cd$^{-1}$. Three of them have a half-amplitude larger than 0.04 mag;
harmonic and combination terms are also observed. The latter result ruled
out the simple model of a binary star composed of two HADSs. The strong
changes in the shape are the largest observed in the amplitude of a HADS
(Poretti 2000). Maybe V974 Oph can be considered as a link between the
multiperiodic low amplitude $\delta$~Sct stars and the stable fundamental
radial HADS pulsators.

\section{Summing-up and Conclusions}
In our travel through the observational scenario we met many objects which
teach us something about the pulsation of $\delta$ Sct stars. The high
frequencies shown by XX Pyx, the intermediate frequencies displayed by FG Vir
and the low frequencies found in the case of
4 CVn  are leading toward the idea that the mode excitation is really
complex and unpredictable. In this respect, the close similarity between 
4 CVn and HD~2724 is comfortable. Probably the multiperiodic behaviour
of the high-amplitude $\delta$ Sct star V974 Oph can also be seen as a 
simplification of the phenomenology, demonstrating that we can observe a 
multimode excitation even when  a large pulsational energy is involved.
To complete the similarities in the opposite direction,
we observe monoperiodic pulsators with a very small amplitude 
(HD 19279 and $\beta$ Cas). What the excited mode is should be
 investigated in order to
understand if a  selection effect acts for these low-amplitude monoperiodic
stars and what difference there is with a star such as AZ CMi, which displays an
asymmetrical light curve.

The amplitude variations are observed in a large variety of stars, both
multiperiodic (XX Pyx, 4 CVn, 44 Tau) and monoperiodic (28 And and BF
Phe).  The observations of mode growth in the light curve of a relatively
simple pulsator such as V663 Cas clarifies what can happen in much more
complicated ones: some modes can be damped and then re-excited. There
are some observational facts which make this explanation preferable even
if the model of a beating between two close frequencies with similar,
constant amplitude cannot be ruled out.  In this context the continuous
survey of $\delta$ Sct stars which will be performed by space missions
could greatly improve the situation; in the case of intrinsic variations,
the time-scale of such variations could be determined.  \smallskip

The better focusing of the phenomenological scenario is not the only
result.  The effort made by the observers in the last years allowed us
to detect a large number of frequencies in stars such as FG Vir, XX Pyx,
4 CVn and BH Psc. By means of multisite campaigns we are able to detect
terms with amplitudes less than 1 mmag; in these conditions ground-based
observations can be successfully complementary to space missions. The
mode identification techniques are based on both the phase shifts and
amplitude ratios of the light and colour curves and the synergic approach
performed by considering spectroscopic curves. Their full exploitation
should guarantee an important role for our researche in stellar physics.

\section{The future: the exportation of the results on galactic stars to
extragalactic research}

\begin{figure}[ht]
\setlength{\textwidth}{4.3in}
\plotone{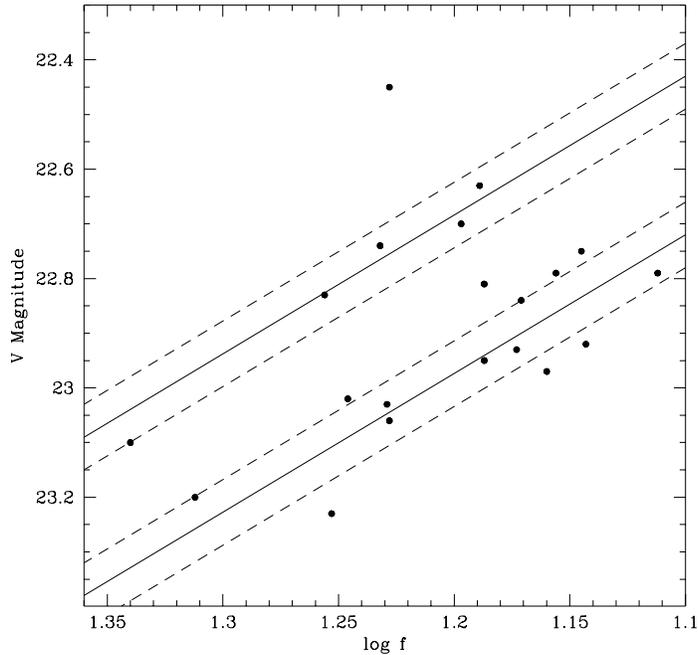}
\caption{The short-period variable stars in the Carina dwarf Spheroidal
galaxy. The Period--Luminosity--Metallicity relationships for fundamental
mode (lower line) and first overtone (upper line) pulsators are shown together
with the frequency values. Dashed lines indicate error bars.}
\label{carina}
\end{figure}

There is always a bit of confusion about the taxonomy of short-period
pulsating stars: in the Galaxy they are called $\delta$ Sct stars if
Pop.~I objects and SX Phe stars if Pop.~II objects.  The old definitions
``dwarf Cepheids'' or RRs or AI Vel stars are currently avoided in the
dedicated literature, but they turn up again in papers on extragalactic
researches. However, the definition currently used by stellar researchers,
i.e., high-amplitude $\delta$ Sct stars (HADS), seems to be irrespective
of which Population they belong to and we are coming back to old
definitions with new names. Moreover, the phenomenology previously
described warns us that the separation between high-- and low-amplitude
$\delta$ Sct stars is not so obvious and that the amplitude cannot be
considered a good physical criterion to separate variable stars.

Mateo, Hurley-Keller, \& Nemec (1998) reported on the discovery of 20
pulsating stars in the
Carina dwarf Spheroidal Galaxy. Their periods range  from 0.048~d to
0.077~d and their amplitude is below 0.30 mag in $V$-light.
Poretti (1999) improved the period values first determined by Mateo et al.
(1998), obtaining a more clear picture of the Period--Luminosity--Metallicity
relationships (Fig.~\ref{carina}).

The observed stars are monoperiodic and  their periods are usually shorter
than the periods observed in galactic HADS. An extended effort was made
to understand the properties of the Fourier parameters of galactic HADS
(Morgan et al.  1998), particularly to use them as a mode discriminant. In
the case of an external galaxy the approach to the analysis of HADS
light curves is the opposite: the Period--Luminosity relationships
allow us to separate in a straightforward way the pulsation modes. Then
we could have the tool to investigate the differences thus driven,
always using the
Fourier decomposition.  In this direction,  the results described by
Poretti (1999) are quite preliminary, but we can be confident that in
the near future we will be able to expand the horizon of HADS studies
to extragalactic topics.

\acknowledgments
The author wishes to thank M.~Breger, G.~Handler, M.~Papar\'o,
E.~Rodr\`{\i}guez, and C.~Sterken for comments on the first draft,
L.~Mantegazza for useful discussions, and J.~Vialle for checking the
English form.

\end{document}